\documentclass[epj]{svjour}

\def\be{\begin{equation}} 
\def\ee{\end{equation}}   

\RequirePackage{graphicx}

\usepackage{cite} 
\usepackage{amsmath}
\usepackage{hyperref}
\usepackage{cite}
\usepackage{amsmath,amssymb}
\usepackage{xcolor}

\usepackage{graphics}

\usepackage{cite} 
\usepackage{amsmath}
\usepackage{hyperref}
\usepackage{cite}
\usepackage{amsmath,amssymb}
\usepackage{xcolor}

\begin{document}
\title{Minimal Geometric Deformation in a cloud of strings}

\author{
Grigoris Panotopoulos \inst{1} 
\thanks{E-mail: \href{mailto:grigorios.panotopoulos@tecnico.ulisboa.pt}{\nolinkurl{grigorios.panotopoulos@tecnico.ulisboa.pt}} }
\and
\'Angel Rinc\'on \inst{2}
\thanks{E-mail: \href{mailto:arrincon@uc.cl}{\nolinkurl{arrincon@uc.cl}} }
}                     
%
%
\institute{ 
Centro de Astrof\'{\i}sica e Gravita{\c c}{\~a}o, Departamento de F{\'i}sica, Instituto Superior T\'ecnico-IST,
\\
Universidade de Lisboa-UL, Av. Rovisco Pais, 1049-001 Lisboa, Portugal
\and
Instituto de F{\'i}sica, Pontificia Cat{\'o}lica Universidad de Chile,
Av. Vicu{\~n}a Mackenna 4860, Santiago, Chile
}
\date{Received: date / Revised version: date}
%
\abstract{
We find new exact analytical solutions in three-dimensional gravity 
applying the Minimal Geometric Deformation approach in a cloud of strings.
\PACS{
      {PACS-key}{discribing text of that key}   \and
      {PACS-key}{discribing text of that key}
     } 
} 
\maketitle
%

\section{Introduction}

Einstein's General Relativity (GR) \cite{GR}, is nowadays considered to be one of the cornerstones of modern theoretical physics, and it provides us with the
framework to adequately describe and understand various aspects of astrophysical
objects and Cosmology. Many of its predictions have been verified observationally, starting from the classical tests in the old days \cite{weinberg,misner}, and recently with the historical LIGO's direct detection of gravitational waves from black holes mergers \cite{ligo1,ligo2,ligo3}. For a recent review on the tests of GR see \cite{tests}.

In spite of its mathematical beauty, handling problems of physical relevance in GR is usually a formidable task. Since it is a highly non-linear theory, the principle of superposition valid in linear differential equations does not apply here, and finding exact solutions has been always a challenge,
see \cite{solutions} for known exact solutions to Einstein's field equations.

A new elegant method that allows us to obtain new exact solutions starting from a known one has received considerable attention recently \cite{mgd}. The so-called Minimal Geometric Deformation (MGD) approach, which was originally introduced in \cite{Ovalle:2007bn} in the context of the brane-world scenario \cite{Randall:1999ee,Randall:1999vf}, has been proven to be a powerful tool in the investigation of the properties of self-gravitating objects, such as relativistic stars \cite{Gabbanelli:2018bhs,Estrada:2018zbh,Heras:2018cpz,Morales:2018nmq,Sharif:2018pzr,Morales:2018urp,Estrada:2018vrl} 
or black hole solutions 
\cite{Ovalle:2018umz,Casadio:2015gea,Ovalle:2015nfa}, see also \cite{Ovalle:2008se,Ovalle:2010zc,Casadio:2012pu,Casadio:2012rf,Ovalle:2013xla,Ovalle:2014uwa}.

The MGD approach has been successfully extended in \cite{Casadio:2015gea} and applied in \cite{Fernandes-Silva:2018abr}, highlighting the potential and power of this new technique. More recently, a method to obtain the isotropic generator of any anisotropic solution was developed in \cite{Contreras:2018gzd} and applied in \cite{Contreras:2018nfg}.

The Ba{\~n}ados, Teitelboim and Zanelli (BTZ) black hole solution \cite{BTZ1,BTZ2,BTZ3} in (1+2) dimensions marked the birth of the interest in lower-dimensional gravity. The absence of propagating degrees of freedom as well as its deep connection to the Chern-Simons term only \cite{chern1,chern2,chern3} make three-dimensional gravity special, and at the same time a framework which allow us to get insight into realistic black holes in four dimensions by studying a mathematically simpler three-dimensional system.

The BTZ black hole is sourced by a negative cosmological constant, but other possibilities, such as scalar fields \cite{mann} and electromagnetic fields \cite{EM1,EM2,Cataldo:2000we} (for studies on the scale-dependent version of some models see \cite{Koch:2016uso,Rincon:2017ypd,Rincon:2017goj,Rincon:2018lyd,Rincon:2018dsq,Rincon:2017ayr,Rincon:2018sgd}) also exist. One option less studied in the literature, which leads to a black hole solution alternative to the BTZ one, is a cloud of strings \cite{letelier}. The matter contribution is described by the Nambu-Goto action, which is well-known both from string theory \cite{polchinski} and from the study of topological defects \cite{vilenkin}. The black hole solution was obtained in \cite{cloud1}, and for related studies on the topic see e.g. \cite{cloud2,cloud3,cloud4,cloud5}.

In the present work we apply the MGD approach to obtain new exact solutions in (1+2) gravity, starting from the known solution (for which the coupling constant $\alpha=0$, see next section) that corresponds to a cloud of strings. We follow closely a recent work \cite{ernesto}, in which the authors applied the MGD approach and obtained new exact solutions in (1+2) gravity, where the known solution corresponded to the BTZ one. Our work is organized as follows: In the next section we briefly present the MGD method, and we apply it to obtain new solutions in the third section. Finally we conclude in section 4. We adopt the mostly negative metric signature, $(+,-,-)$.

\section{Field equations and Minimal Geometric Deformation}

The starting point is Einstein's field equations
\be
G_{\mu \nu} = R_{\mu \nu} - \frac{1}{2} R g_{\mu \nu}  = - \kappa^2 T_{\mu \nu}
\ee
where $\kappa^2 = 8 \pi G$, and the total stress-energy tensor $T_{\mu \nu}$ has two contributions
\be
T_{\mu \nu} = M_{\mu \nu} + \alpha \theta_{\mu \nu}
\ee
The first source $M_{\mu \nu}$ is supposed to lead to a known solution, while the second source is coupled to the first one via the coupling constant $\alpha$.

Seeking static circularly symmetric solutions we adopt the coordinate system $(t,r,\phi)$, we make as usual for the metric tensor the ansatz
\be
\mathrm{d}s^2 = e^{\nu(r)} \mathrm{d}t^2 - e^{\lambda(r)} \mathrm{d}r^2 - r^2 \mathrm{d} \phi^2
\ee
and we obtain the following set of coupled differential equations for the two unknown metric functions $\nu(r),\lambda(r)$
\begin{eqnarray}
\kappa^2 (M_0^0 + \alpha \theta_0^0) & = & - \frac{\lambda' e^{-\lambda}}{2 r}  \\
 \kappa^2 (M_1^1 + \alpha \theta_1^1) & = &  \frac{\nu' e^{-\lambda}}{2 r} \\
 \kappa^2 (M_2^2 + \alpha \theta_2^2) & = &  - \frac{e^{-\lambda}}{4} [\nu' (\lambda'-\nu')-2 \nu'']
\end{eqnarray}
where the prime denotes differentiation with respect to the radial coordinate.

Notice that the system at hand may be viewed as an anisotropic fluid with energy momentum tensor
\be
T_\nu^\mu = \text{diag}(\rho, -p_r, -p_t) 
\ee
where the energy density $\rho$ and the pressures $p_r,p_t$ are given by
\begin{align}
\rho &\equiv M_0^0 + \alpha \theta_0^0 \label{rho}
\\
p_r &\equiv -(M_1^1 + \alpha \theta_1^1) \label{p1}
\\
p_t &\equiv -(M_2^2 + \alpha \theta_2^2) \label{p2}
\end{align}
while the anisotropy $\Delta \equiv p_t-p_r$ is given by
\be
\Delta = M_1^1 - M_2^2 + \alpha (\theta_1^1 - \theta_2^2) \label{anisotropy}
\ee
It is worth mentioning that in the three-dimensional case studied here both the isotropic and the anisotropic sector satisfy Einstein's equations, contrary to the four-dimensional cases where the anisotropic sector satisfies "quasi-Einstein" field equations \cite{mgd,Ovalle:2018umz}. This 
was already pointed out in \cite{ernesto}.

The MGD method allows us to solve the full problem in two steps as follows: First, we assume that the solution to the simpler problem for $\alpha=0$ is known, for some given functions $\nu(r)$ and $e^{- \lambda(r)}=\mu(r)$. Then, when we turn the $\alpha$ parameter on we assume for simplicity that the first metric function $\nu(r)$ remains the same, while the presence of the second source modifies the second metric function as follows
\be
e^{- \lambda(r)} = \mu(r) + \alpha h(r)
\ee
where the so called deformation function satisfies the equations
\begin{eqnarray}
2 \kappa^2 r \theta_0^0 & = &  h'  \\
2 \kappa^2 r \theta_1^1 & = &  h \nu'  \\
4 \kappa^2 \theta_2^2 & = &  [ h' \nu' + 2 h \nu'' + h (\nu')^2 ]
\end{eqnarray}
Finally, since the two energy momentum tensors are separately conserved, the components of the second source must satisfy the following condition
\be
(\theta_1^1)' - \frac{1}{2}\nu'(\theta_0^0 - \theta_1^1) - \frac{1}{r}(\theta_2^2-\theta_1^1) = 0 \label{conservation}
\ee

\section{New exact solution in 3D cloud of strings}

Here, following \cite{ernesto}, we apply the MGD method to obtain exact analytical solutions to three-dimensional Einstein's field equations, where some source with stress-energy tensor $\theta_{\mu \nu}$ is coupled to a cloud of strings with stress-energy tensor 
\be
M_\nu^\mu = \frac{\xi}{r} \text{diag}(1,1,0)
\ee
so we may identify the corresponding fluid parameters to be
\begin{align}
\rho^{\text{cloud}}  & =  \frac{\xi}{r}
\\
p_r ^{\text{cloud}}  & = -\frac{\xi}{r}
\\
p_t^{\text{cloud}}   & =  0
\end{align}

The black hole solution corresponding to this kind of matter was obtained in \cite{cloud1}, and it is given by
\be
e^{\nu(r)} = \mu(r) = -M + 2 \xi r
\ee
with $M$ being the mass of the black hole, and there is a single horizon at $r_H=M/(2 \xi)$.
This would be the known solution corresponding to the initial simple problem for $\alpha=0$. To obtain the full solution when both terms in the stress-energy tensor are present we have to determine the deformation function.

\subsection{General constraint}

Since there are four unknown functions (the deformation and the components of the second gravitational source) and only three independent equations, it is necessary to assume a certain condition between the components of $\theta_\nu^\mu$ in order to close the system of equations and
obtain the solution. Therefore, similar to \cite{ernesto}, here we shall impose the constraint
\be
\theta_1^1 = a \theta_0^0 + b \theta_2^2
\ee
with two arbitrary constant parameters $a,b$. Using the equations satisfied by the deformation function we obtain an ordinary differential equation of first order for $h(r)$ of the form
\be
\frac{\mathrm{d}h}{\mathrm{d}r} = \frac{B}{A} h
\ee
where the functions $A(r),B(r)$ are found to be
\begin{eqnarray}
A & = & \frac{a}{r} + \frac{b \nu'}{2}  \\
B & = & \frac{\nu'}{r} -b \nu'' - \frac{b (\nu')^2}{2}
\end{eqnarray}
The equation above can be integrated directly and we obtain for the deformation function the expression
\be
h = c_1 \frac{-M + 2 \xi r}{|-aM+\xi(2a+b)r|^k}
\ee
where $c_1$ is an arbitrary integration constant, while the power $k$ is given by
\be
k = \frac{2(a-1)}{2a+b}
\ee
In the following subsections we consider two concrete examples.

\subsection{Particular constraint $\#$ 1}

First, let us assume that $\theta_1^1 = \theta_2^2$, which corresponds to $a=0,b=1$. In this case the deformation function takes the form
\be
h(r) = c_1 (\xi r)^2 (-M + 2 \xi r)
\ee
and therefore the metric function is computed to be
\begin{align}
e^{-\lambda} &= \Bigl[1 + \alpha c_1  \bigl( \xi r \bigl)^2 \Bigl] (2 \xi  r-M)
\end{align}

The components of $\theta_{\mu \nu}$ can be explicitly computed one by one
\begin{align}
\theta_1^1 &= \theta_2^2 =  \frac{c_1 \xi^3}{\kappa^2}r
\\
\theta_0^0 &=  \frac{c_1 \xi^2}{\kappa^2} (-M + 3 \xi r)
\end{align}
and it is easy to verify that for the solution just obtained the condition of energy conservation (\ref{conservation}) is satisfied.

The fluid parameters can be computed using the equations (\ref{rho}), (\ref{p1}) and (\ref{p2}) to obtain
\begin{align}
\rho &=  \frac{\xi }{r} - \alpha \ \frac{c_1 \xi ^2 }{\kappa ^2}(M - 3 \xi  r) 
 \\
p_r  &= -\frac{\xi }{r} -  \alpha \ \frac{c_1 \xi ^3 }{\kappa ^2} r
\\
p_t  &=                 -  \alpha \ \frac{c_1 \xi ^3 }{\kappa ^2} r
\end{align}
and the anisotropy may be easily computed using its definition $\Delta = p_t-p_r$.

Finally, to check for potential singularities we compute the Ricci scalar as well as the Kretschmann scalar, which are found to be
\begin{align}
R = -\frac{4 \xi}{r}  &\bigg[1-\frac{3}{2} \alpha  c_1 \xi  r (M-4 \xi  r) \bigg]
\\  
\begin{split}
K = \frac{8 \xi^2}{r^2} 
&\bigg[
1 
-
2 \alpha  c_1 \xi  r (M-3 \xi  r)
\ +
\\
& \ \frac{3}{2} \alpha ^2 c_1^2 \bigl(\xi r \bigl)^2 \left(M^2-8 M \xi  r+18 \xi ^2 r^2\right)`
\bigg]
\end{split}
\end{align}
Clearly, in both expressions the first term comes from the cloud of strings, while the other terms come from the coupling between the sources. The only singularity is the usual singularity at the center as $r \rightarrow 0$.

\begin{figure*}[ht]
\centering
\includegraphics[width=0.32\textwidth]{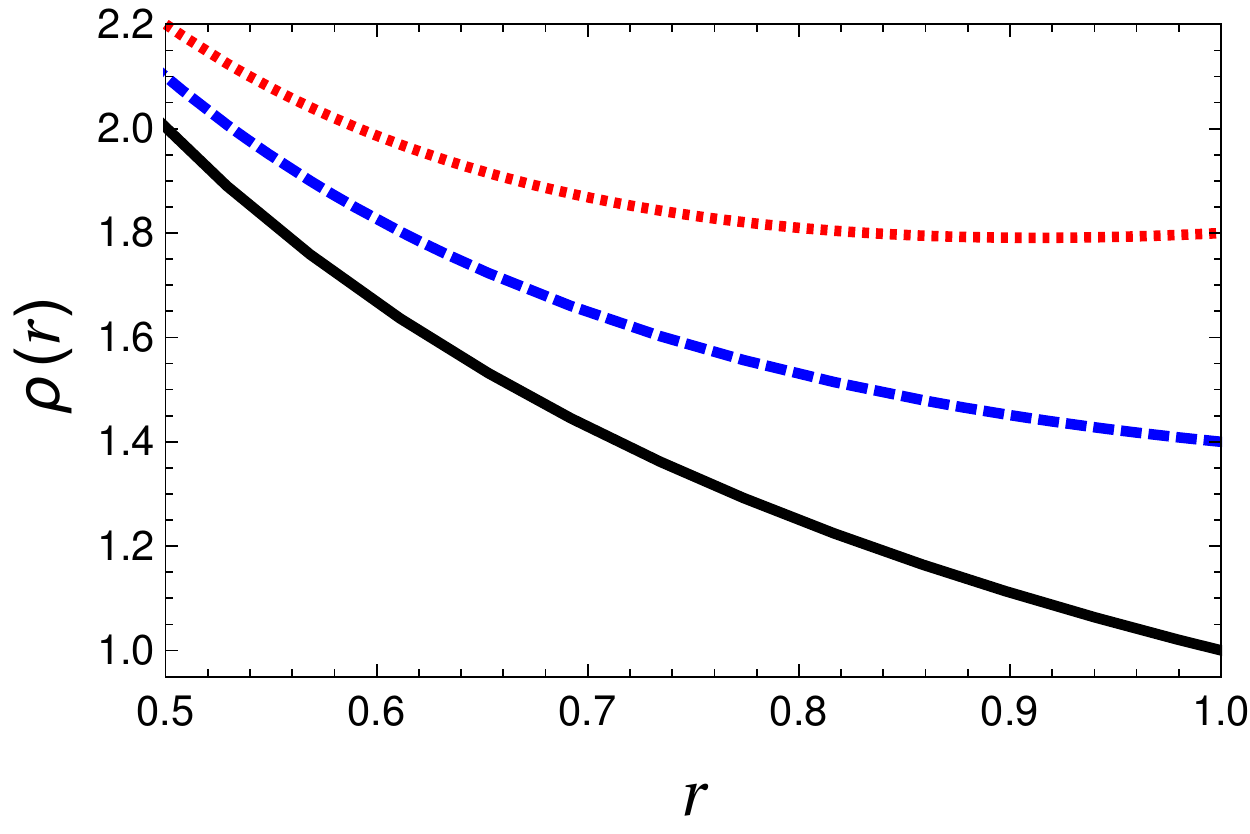}     \
\includegraphics[width=0.32\textwidth]{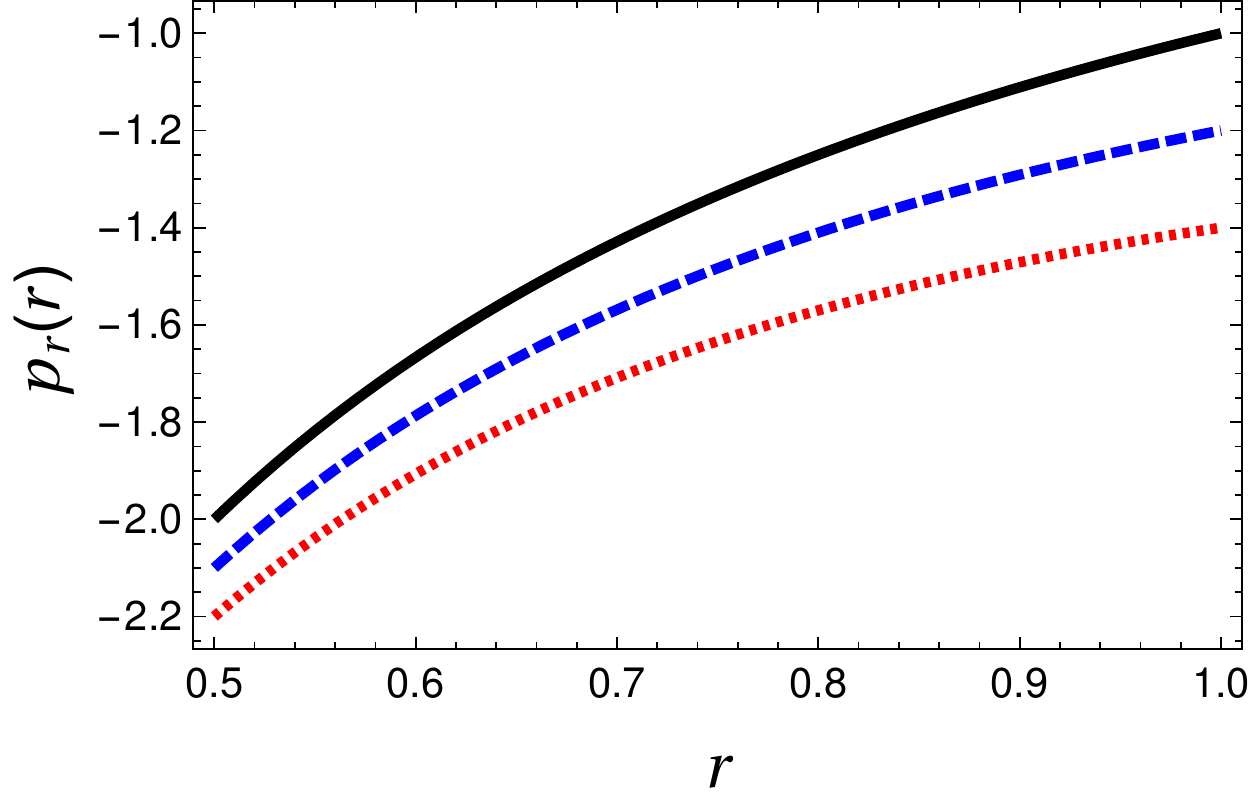}      \
\includegraphics[width=0.32\textwidth]{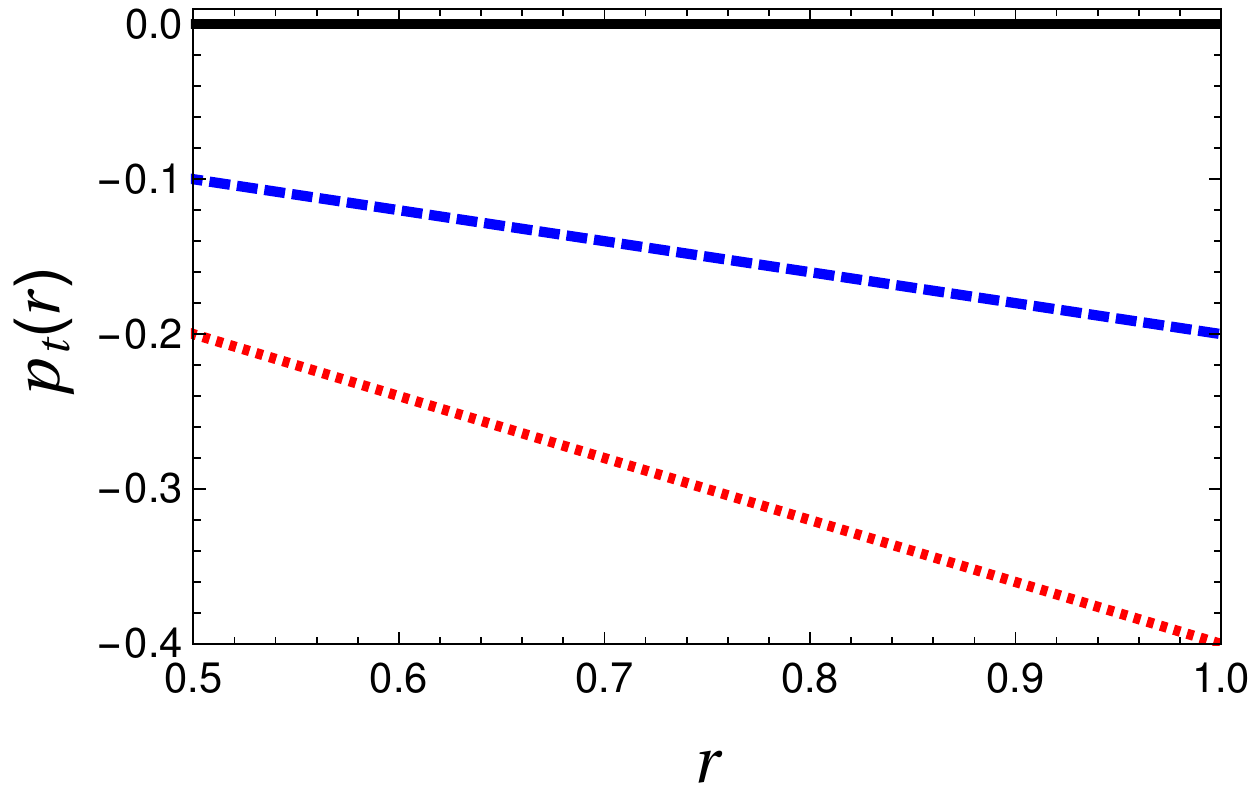}      \
\caption{
Fluid parameters ${\rho, p_r, p_t}$ versus radial coordinate $r$ for the particular constraint $\#$ 1.
{\bf Left panel:} Energy density $\rho$ vs radial coordinate $r$ for different values of the parameter $\alpha$.  {\bf Middle panel:}
Radial pressure $p_r$ vs radial coordinate $r$ for different values of the parameter $\alpha$.
{\bf Right panel:} Tangential pressure $p_t$ vs radial coordinate $r$ for different values of the parameter $\alpha$. Shown are: i) $\alpha = 0$ (solid black line), ii) $\alpha = 0.2$ (dashed blue line) and, iii) $\alpha = 0.4$ (dotted red line). The other two parameters have been taken equal to unity, $M=\xi=1$.
}
\label{fig:2}
\end{figure*}

\subsection{Particular constraint $\#$ 2}

In the second example let us assume that $\theta_{\mu \nu}$ is traceless, $\theta_\mu^\mu=0$,
which corresponds to the case $a=-1=b$. In this case the deformation function takes the form
\be
h(r) = c_1 \frac{-M + 2 \xi r}{(-M + 3 \xi r)^{4/3}}
\ee
and, as before, the metric function can be written in terms of the deformation function as
\begin{align}
e^{-\lambda} &= \left[1 + \alpha c_1  \Bigg(\frac{1}{3 \xi  r - M} \Bigg)^{4/3}\right]
 (2 \xi  r-M)
\end{align}
while the components of $\theta_{\mu \nu}$ are computed to be
\begin{eqnarray}
\theta_0^0 & = & - c_1 \frac{\xi (-M + \xi r)}{\kappa^2 r (-M + 3 \xi r)^{7/3}}  \\
\theta_1^1 & = &  c_1 \frac{\xi}{\kappa^2 r (-M + 3 \xi r)^{4/3}} \\
\theta_2^2 & = & - c_1 \frac{2 \xi^2}{\kappa^2 (-M + 3 \xi r)^{7/3}}
\end{eqnarray}
It is easy to check that the trace is indeed zero.
The fluid parameters can be computed using the equations (\ref{rho}), (\ref{p1}) and (\ref{p2}) to obtain
\begin{align}
\rho &= \frac{\xi }{r} - \alpha \ \frac{c_1 \xi}{\kappa ^2 r} \frac{ (\xi  r-M)}{(3 \xi  r-M)^{7/3}}
\\
p_r  &= -\frac{\xi }{r} - \alpha \ \frac{c_1 \xi}{\kappa ^2 r} 
\Bigg(\frac{1}{3 \xi  r - M} \Bigg)^{4/3}
\\
p_t  &=   \alpha \ \frac{2 c_1 \xi ^2}{\kappa ^2 (3 \xi  r-M)^{7/3}}
\end{align}
and the anisotropy may be easily computed using its definition $\Delta = p_t-p_r$.

Similarly to the previous case, to check for potential singularities we compute the Ricci scalar as well as the Kretschmann scalar, which are computed to be
\begin{align}
R &=  - \frac{4 \xi}{r} 
\Bigg[
1 - \alpha c_1 \frac{\left(M^2-M \xi  r+\xi ^2 r^2\right)}{(3 \xi  r-M)^{10/3}}
\Bigg]
\\  
\begin{split}
K &= \frac{8 \xi ^2}{r^2}
\Bigg[
1 
+ 
\alpha  c_1 \frac{2 (M-\xi  r)}{(3 \xi  r-M)^{7/3}} 
+
\bigl(\alpha c_1\bigl)^2 \ \times
\\
&
\frac{2 \xi ^2 r^2 (3 M-2 \xi  r)^2+(M-\xi  r)^2 (M-3 \xi  r)^2}{(-M + 3 \xi r)^{20/3}}
\Bigg]
\end{split}
\end{align}
We recover the expressions corresponding to a cloud of strings in the limit $\alpha \rightarrow 0$.
We see that apart from the usual singularity at $r=0$, there is another one 
at $r_{\star} = M/(3 \xi) < r_H$. Therefore we conclude that since it is located into the forbidden zone it is not a physical singularity.

We remark in passing that in both concrete examples considered here the horizon remains the same. In the figures we show the impact of the coupling constant $\alpha$ on the solution in the two concrete examples setting $c_1=M=\xi=1$.

\begin{figure*}[ht]
\centering
\includegraphics[width=0.32\textwidth]{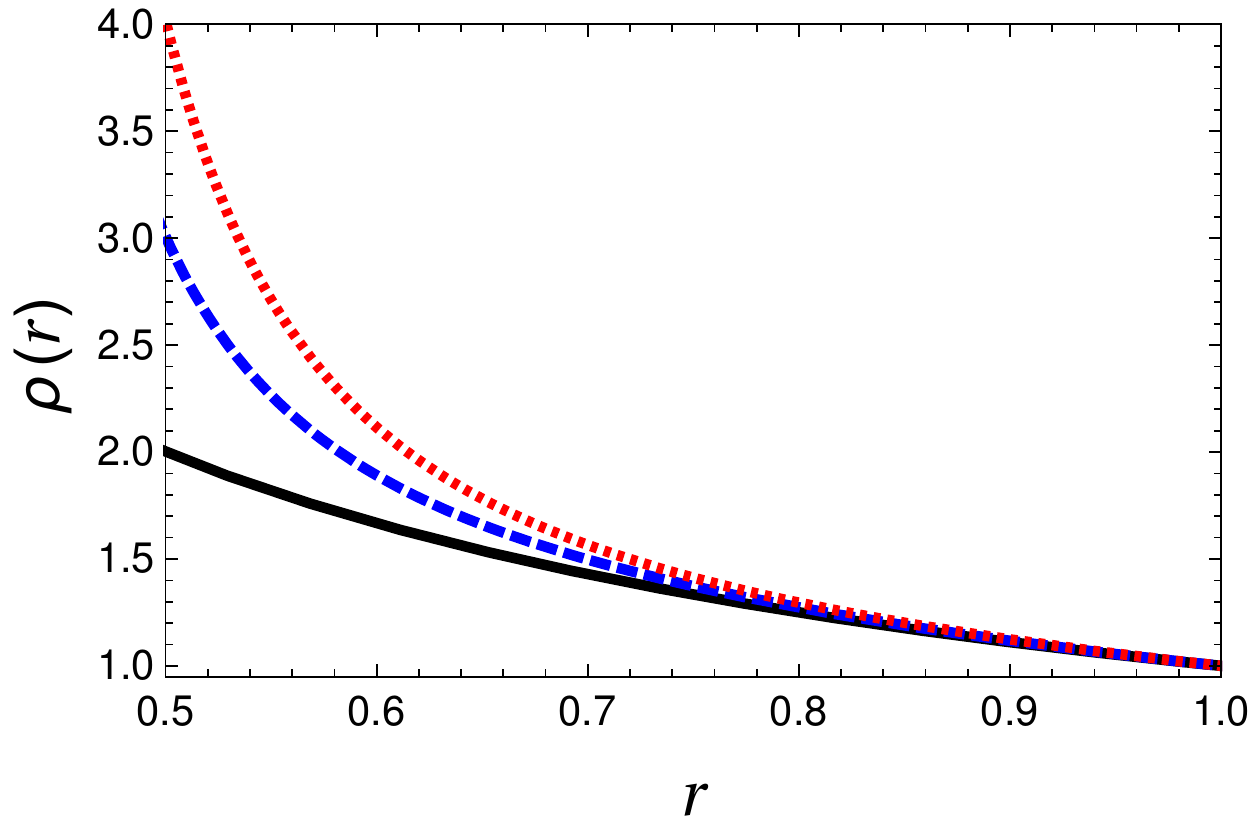}     \
\includegraphics[width=0.32\textwidth]{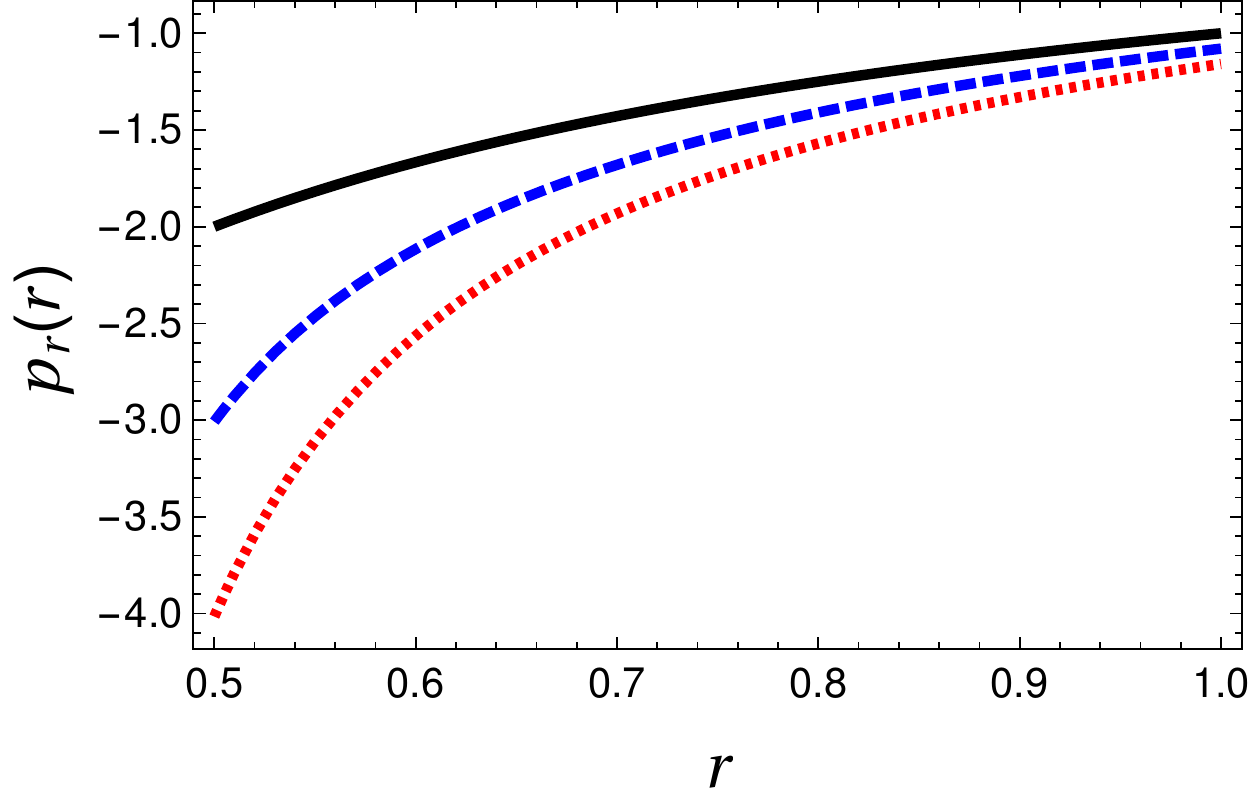}      \
\includegraphics[width=0.32\textwidth]{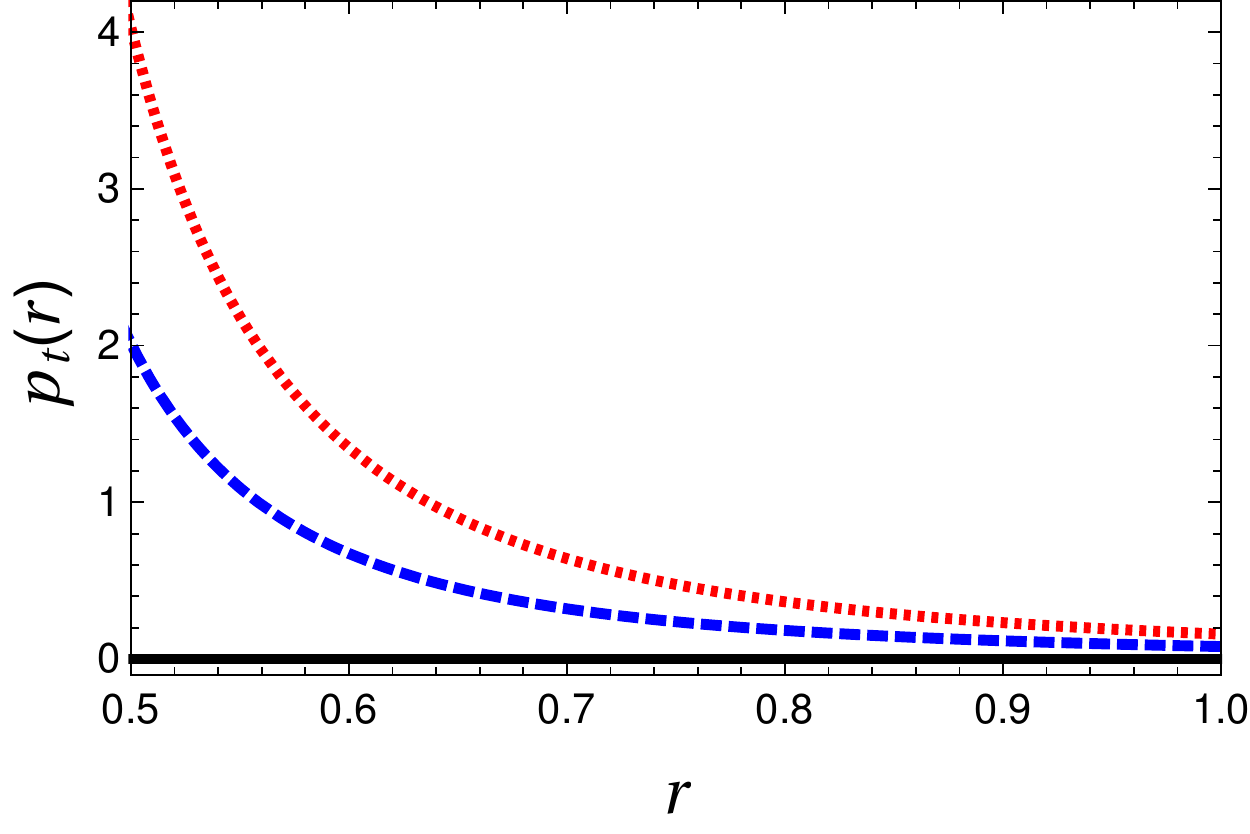}      \
\caption{
Fluid parameters ${\rho, p_r, p_t}$ versus radial coordinate $r$ for the particular constraint $\#$ 2.
{\bf Left panel:} Energy density $\tilde{\rho}$ vs radial coordinate $r$ for different values of the parameter $\alpha$.  {\bf Middle panel:}
Radial pressure $p_r$ vs radial coordinate $r$ for different values of the parameter $\alpha$.
{\bf Right panel:} Tangential pressure $p_t$ vs radial coordinate $r$ for different values of the parameter $\alpha$. Shown are: i) $\alpha = 0$ (solid black line), ii) $\alpha = 0.2$ (dashed blue line) and, iii) $\alpha = 0.4$ (dotted red line). The other two parameters have been taken equal to unity, $M=\xi=1$.
}
\label{fig:3}
\end{figure*}

\begin{figure*}[ht]
\centering
\includegraphics[width=0.48\textwidth]{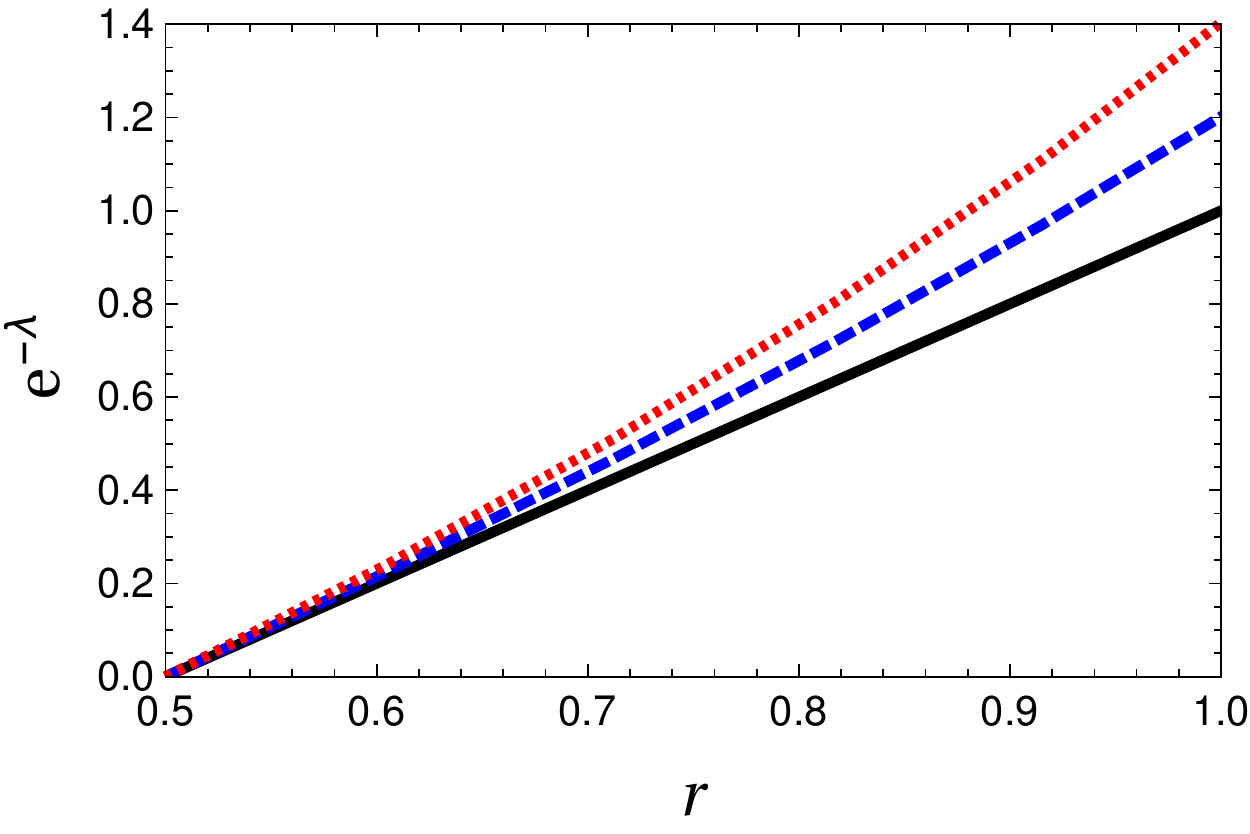}     \
\includegraphics[width=0.48\textwidth]{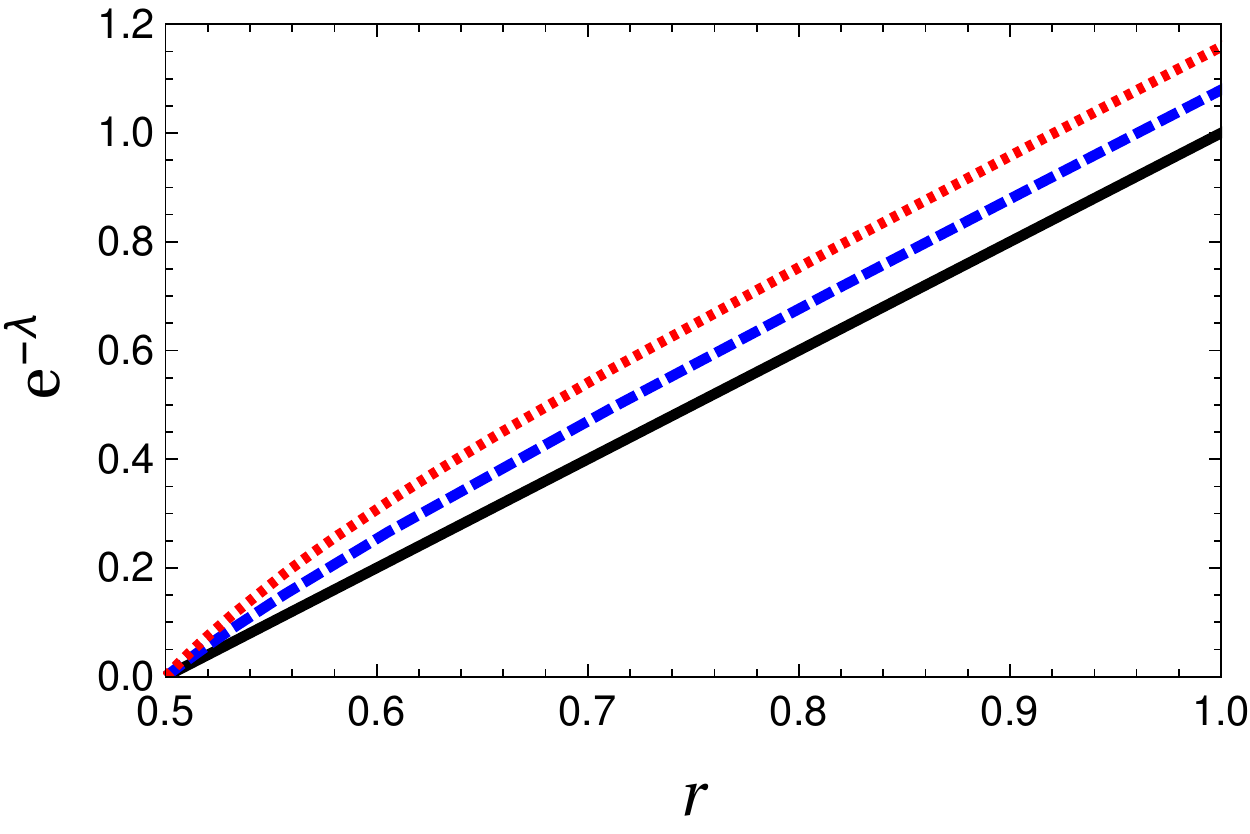}      
\caption{
{\bf Left panel:} Metric function $\text{e}^{-\lambda}$ vs radial coordinate $r$ for the particular constraint $\#$ 1 assuming different values of the parameter $\alpha$.  {\bf Right panel:}
Metric function $\text{e}^{-\lambda}$ vs radial coordinate $r$ for different values of the parameter $\alpha$ for the particular constraint $\#$ 2 assuming different values of the parameter $\alpha$. Shown are: i) $\alpha = 0$ (solid black line), ii) $\alpha = 0.2$ (dashed blue line) and, iii) $\alpha = 0.4$ (dotted red line). The other two parameters have been taken equal to unity, $M=\xi=1$.
}
\label{fig:3}
\end{figure*}

\section{Conclusions}

To summarize, in the present work we have obtained new exact analytical solutions in a three-dimensional cloud of strings applying the Minimal Geometric Deformation approach.
Two concrete examples are presented in detail, where the Ricci and Kretschmann scalars are computed too, and the impact of the coupling constant on the solution is investigated. We find, among other things, that there is a single horizon and the usual singularity at the center (and no other) with or without the additional source.


\section*{Acknowlegements}
We wish to thank the anonymous reviewer for useful comments and suggestions. The work of A.R. was supported by the CONICYT-PCHA/Doctorado Nacional/2015-21151658. 
G.P. thanks the Funda\c c\~ao para a Ci\^encia e Tecnologia (FCT), Portugal, for the financial support to the Center for Astrophysics and Gravitation-CENTRA, Instituto 
Superior T\'ecnico, Universidade de Lisboa, through the Grant No. UID/FIS/00099/2013.


\end{document}